\documentclass[aps,prl,twocolumn,superscriptaddress,floatfix]{revtex4-1}
\usepackage[colorlinks=true,citecolor=blue,linkcolor=blue,urlcolor=blue]{hyperref}
\usepackage{graphicx, nicefrac, textcomp}
\usepackage{hyperref}

\begin{document}

\title{Control of the metal-insulator transition in NdNiO$_\mathbf{3}$ thin films through the interplay between structural and electronic properties}

\author{Y.~E.~Suyolcu}
\affiliation{Max Planck Institute for Solid State Research, Heisenbergstrasse 1, 70569 Stuttgart, Germany}
\affiliation{Department of Materials Science and Engineering, Cornell University, Ithaca, New York 14853, USA}

\author{K.~F{\"u}rsich}
\affiliation{Max Planck Institute for Solid State Research, Heisenbergstrasse 1, 70569 Stuttgart, Germany}

\author{M.~Hepting}
\affiliation{Max Planck Institute for Solid State Research, Heisenbergstrasse 1, 70569 Stuttgart, Germany}

\author{Z.~Zhong}
\affiliation{Max Planck Institute for Solid State Research, Heisenbergstrasse 1, 70569 Stuttgart, Germany}
\affiliation{Key Laboratory of Magnetic Materials and Devices and Zhejiang Province Key Laboratory of Magnetic Materials and Application Technology, Ningbo Institute of Materials Technology and Engineering (NIMTE),Chinese Academy of Sciences, Ningbo 315201, China}

\author{Y.~Lu}
\affiliation{Max Planck Institute for Solid State Research, Heisenbergstrasse 1, 70569 Stuttgart, Germany}

\author{Y.~Wang}
\affiliation{Max Planck Institute for Solid State Research, Heisenbergstrasse 1, 70569 Stuttgart, Germany}

\author{G.~Christiani}
\affiliation{Max Planck Institute for Solid State Research, Heisenbergstrasse 1, 70569 Stuttgart, Germany}

\author{G.~Logvenov}
\affiliation{Max Planck Institute for Solid State Research, Heisenbergstrasse 1, 70569 Stuttgart, Germany}

\author{P.~Hansmann}
\affiliation{Max Planck Institute for Solid State Research, Heisenbergstrasse 1, 70569 Stuttgart, Germany}
\affiliation{Max Planck Institute for Chemical Physics of Solids, N{\"o}thnitzer-Strasse 40, 01187 Dresden, Germany}
\affiliation{Department of Physics, University of Erlangen-N\"{u}rnberg, 91058, Erlangen, Germany}

\author{M.~Minola}
\affiliation{Max Planck Institute for Solid State Research, Heisenbergstrasse 1, 70569 Stuttgart, Germany}

\author{B.~Keimer}
\affiliation{Max Planck Institute for Solid State Research, Heisenbergstrasse 1, 70569 Stuttgart, Germany}

\author{P.~A.~van~Aken}
\affiliation{Max Planck Institute for Solid State Research, Heisenbergstrasse 1, 70569 Stuttgart, Germany}

\author{E.~Benckiser}
\email{E.Benckiser@fkf.mpg.de}
\affiliation{Max Planck Institute for Solid State Research, Heisenbergstrasse 1, 70569 Stuttgart, Germany}

\date{\today}

\begin{abstract}

Heteroepitaxy offers a new type of control mechanism for the crystal structure, the electronic correlations, and thus the functional properties of transition-metal oxides. Here, we combine electrical transport measurements, high-resolution scanning transmission electron microscopy (STEM), and density functional theory (DFT) to investigate the evolution of the metal-to-insulator transition (MIT) in NdNiO$_3$ films as a function of film thickness and NdGaO$_3$ substrate crystallographic orientation. We find that for two different substrate facets, orthorhombic (101) and (011), modifications of the NiO$_6$ octahedral network are key for tuning the transition temperature $T_{\text{MIT}}$ over a wide temperature range. A comparison of films of identical thickness reveals that growth on [101]-oriented substrates generally results in a higher $T_{\text{MIT}}$, which can be attributed to an enhanced bond-disproportionation as revealed by the DFT+$U$ calculations, and a tendency of [011]-oriented films to formation of structural defects and stabilization of non-equilibrium phases. Our results provide insights into the structure-property relationship of a correlated electron system and its evolution at microscopic length scales and give new perspectives for the epitaxial control of macroscopic phases in metal-oxide heterostructures.

\end{abstract}
\maketitle

\section{Introduction}

\begin{figure*}[t]
\center\includegraphics[width=0.99\textwidth]{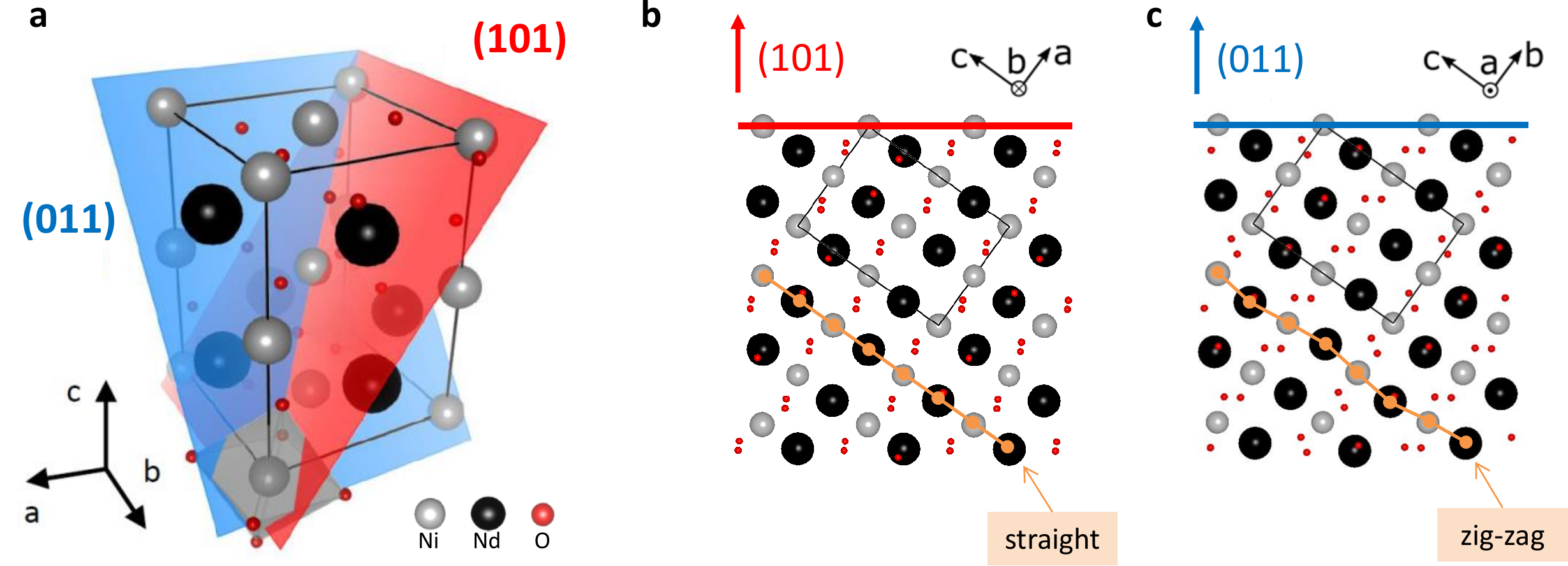}
\caption{(a) Schematic of the orthorhombic NdNiO$_3$ unit cell, with the (101) and (011) crystallographic planes highlighted in red and blue, respectively. Additionally, one NiO$_6$ octahedron is shown in gray. Note that NdGaO$_3$ exhibits a similar unit cell (not shown here) with slightly different lattice parameters (see Methods). (b), (c) Projections of the NdNiO$_3$ crystal structure along the [101] and [011] direction, respectively. The unit cell from panel (a) is indicated by black lines, and the (101) and (011) crystallographic planes by red and blue lines, respectively. The orange lines indicate characteristic \textit{straight} and \textit{zig-zag} patterns of the Ni and Nd cation positions in the two projections. These patterns can be identified in STEM-HAADF images.}
\label{structure}
\end{figure*}

The strong coupling between spin, charge, orbital, and structural degrees of freedom gives rise to exotic ground states in transition-metal oxides (TMOs), such as unusual magnetism, multiferroicity, enhanced thermoelectricity, and high-temperature superconductivity \cite{khomskii2014transition}. The prospect of realizing new, technologically accessible and robust functionality is based on a profound understanding of the mechanisms underlying these phenomena. Along these lines, the structure-property relationship has been considered key in bulk TMOs, yet only few studies have addressed its role in heterostructures. Only recently, the structural modification at heterointerfaces and their relation to the emergence of novel phases has been investigated in detail \cite{Hwang12,zubko2011interface}. This new focus on interfaces is mostly related to tremendous progress in techniques that can obtain detailed structural information about the few atomic layers that constitute an interface. In particular, technical advances in STEM have made it possible to overcome previous limitations, with aberration-corrected STEM not only providing information about the cation sublattice, but also accurately resolving positions of oxygen anions, which are highly relevant for the physics of TMOs \cite{Liao16,Kan16,Dominguez19,Qi15,suyolcu2017octahedral,kim2016polar,suyolcu2017dopant}.

One of the simplest structural motifs of ternary TMOs is the perovskite structure \textit{AB}O$_3$, with usually an alkaline-earth or rare-earth element occupying the \textit{A}-site cation position and a transition-metal element on the \textit{B}-site. The ideal cubic structure is only realized in very few compounds. Most perovskites show distortions, which in first approximation can be described by collective tilts and rotations of rigid \textit{B}O$_6$ octahedra \cite{Glazer72,May10}, resulting in possibly anisotropic \textit{B}-O-\textit{B} bond angles that deviate from the cubic 180$^{\circ}$ ones. This has important consequences for the electronic and magnetic properties, since, for example, electronic bandwidths are reduced such that metal-insulator transitions (MITs) are observed. Furthermore, reduced overlap of TM-$d$ and oxygen-$p$ orbitals can strongly modify the magnetic superexchange and therewith the macroscopic magnetic order. In a heterostructure design, this strong connectivity of structural and electronic properties can be used specifically to stabilize new phases \cite{Chen2021,ramesh2019creating,suyolcu2019design,Hwang12,zubko2011interface,Bluschke17}.

A prototypical TMO with a pronounced structure-property relation is the rare-earth nickelate NdNiO$_3$ \cite{Torrance92, Catalan08, Middey16review, Catalano18}. In bulk form, NdNiO$_3$ exhibits a MIT at approximately 200\,K that coincides with an antiferromagnetic phase transition and a structural phase transition from the orthorhombic space group $Pbnm$ to monoclinic $P2_{1}/n$ \cite{Garica-Munoz09,Lu16}. The MIT is accompanied by a splitting of the uniformly sized NiO$_6$ octahedra into sets of compressed and expanded NiO$_6$ octahedra, corresponding to a rocksalt-pattern breathing-mode distortion. The proposal to consider NdNiO$_3$ as a negative charge-transfer material \cite{Mizokawa00,Subedi15,Green16} has for the first time led to a satisfactory description of experimental studies, in particular from x-ray spectroscopy \cite{Bisogni16,Fuersich19,Hepting14}. In this model the main electronic contribution to the ground state is $3d^8\underline{L}^1$, where $\underline{L}^1$ denotes one oxygen ligand hole. The insulating phase then arises from a ($3d^8\underline{L}^2$, $3d^8$)-type disproportionation of the NiO$_6$ octahedra with modulated Ni-O bond length \cite{Johnston14} rather than a previously proposed charge transfer between adjacent Ni sites \cite{Staub02}.

The electronic and magnetic correlations in NdNiO$_3$ and other rare-earth nickelates are exceptionally sensitive to heteroepitaxial modifications imposed by the substrate \cite{Boris11,Frano13,Middey16review, Catalano18}, which can result in a complete suppression of the MIT for large compressive strain \cite{Hepting14, liu2013heterointerface}. On the other hand, Catalano \textit{et al.} reported that for NdNiO$_3$ films grown on [101]-oriented NdGaO$_3$ substrates, an enhancement of $T_{\text{MIT}}$ by more than 100\,K is possible \cite{Catalano15}. This effect was attributed to the specific three-fold interconnectivity between the film and substrate NiO$_6$-GaO$_6$ octahedral network across the (101) interface (\textit{i.e.} each interfacial Ni forms three bonds via oxygen to Ga ions), together with close lattice matching conditions between NdNiO$_3$ and NdGaO$_3$. These results initiated further studies of films and superlattices on substrates with this unconventional orientation \cite{Middey16,Janson18,kim2016polar,Chakhalian20} that led to important insights into magnetic exchange interactions in thin and ultra-thin NdNiO$_3$ layers \cite{Hepting18,Lu18,Fuersich19}.

To gain a deeper understanding of the role of interfacial bonding on the MIT in nickelates, we have investigated the similarities and differences of NdNiO$_3$ thin films grown on (101) and (011) facets of NdGaO$_3$ substrates, respectively (Fig.~\ref{structure}). For both crystallographic facets, the interconnectivity of the octahedral network across the film-substrate interface is three-fold \cite{Catalano18}. In electrical transport measurements we find that the $T_{\text{MIT}}$ can be systematically varied over a wide temperature range as a function of substrate orientation and film thickness. The microscopic structural information provided by STEM reveals differences in octahedral connectivity, film orientation, and tendency towards stacking fault formation. The experimental findings are corroborated by DFT$+U$ calculations suggesting that energetically disfavored phases can be accommodated by crystallographic stacking faults or a reorientation of the NdNiO$_3$ unit cell along the growth direction. Furthermore, our \textit{ab-initio} calculations reveal an enhanced bond-disproportionation in case of [101]-oriented NdNiO$_3$ under tensile strain, rationalizing the enhanced $T_{\text{MIT}}$ in our [101]-oriented films and those of Ref.~\onlinecite{Catalano15}.

\section{Methods}\label{methods}

Pulsed laser deposition (PLD) was used to grow NdNiO$_3$ thin films on NdGaO$_3$ substrates with (011) and (101) crystallographic surfaces (in $Pbnm$ notation; see Fig.~\ref{structure}), respectively. For each growth, NdNiO$_3$ was deposited simultaneously on a [011]- and [101]-oriented NdGaO$_3$ substrate, yielding pairs of films grown under the same conditions with nominally identical film thicknesses. Details about the growth procedure can be found in Ref.~\onlinecite{Wu13}. The film thicknesses 106, 104, 70, and $40\,\text{\AA}$ were determined from x-ray diffraction and reflectivity measurements. Within our experimental error we obtain identical thicknesses from the analysis of Laue and Kiessig fringes, indicating a single crystalline phase of the whole film. Additionally, an ultra-thin $14.4\,\text{\AA}$ NdNiO$_3$ film was grown on (101) NdGaO$_3$ and capped by a thick NdGaO$_3$ layer.
Both, bulk NdNiO$_3$ and NdGaO$_3$ exhibit a \textit{Pbnm} structure at room-temperature, with lattice parameters $a = 5.387\,\text{\AA}$, $b = 5.383\,\text{\AA}$, $c = 7.610 \,\text{\AA}$, and $a=5.428\,\text{\AA}$, $b = 5.498\,\text{\AA}$, and $c = 7.708\,\text{\AA}$, respectively \cite{Garica-Munoz09, vasylechko2000crystal}. Thus, the larger NdGaO$_3$ unit cell dimensions can impose tensile strain on epitaxially grown NdNiO$_3$. The (011) and (101) NdGaO$_3$ facets correspond to cuts perpendicular to the $[111]_{\text{pc}}$ and $[\bar111]_{\text{pc}}$ body diagonal of the pseudocubic unit cell, respectively. Note that while in the simplified pseudocubic reference frame all body diagonal directions are equivalent, the [011] and [101] direction of the orthorhombic \textit{Pbnm} unit cells are distinct. Projections of the crystal structure perpendicular to the (011) and (101) plane are shown in Figs.~\ref{structure}(b,c), with a characteristic \textit{zig-zag line} connecting the Ni-Nd cation positions in the former projection, and a \textit{straight line} in the latter one. In the following, we will use these distinct line patterns to identify the orientation of the substrate and the film in the STEM high-angle annular dark-field (HAADF) images (Figs.~\ref{8MLTEM}(a), \ref{TEM_KF}).

The electron-transparent specimens were prepared by employing mechanical grinding, tripod-wedge polishing and argon-ion milling steps, respectively. We used a precision-ion polishing system (Gatan PIPS II, Model 695) at low temperature for argon-ion thinning. The STEM investigations were performed with a JEOL JEM-ARM200F microscope equipped with a cold field-emission electron source, a probe Cs-corrector (DCOR, CEOS GmbH), and a Gatan GIF Quantum ERS spectrometer at 200\,kV. STEM imaging and EELS analysis were conducted at probe semi-convergence angles of 20\,mrad and 28\,mrad, resulting in probe sizes of $0.8\,\text{\AA}$ and $1.0\,\text{\AA}$, respectively. Collection angles for annular dark-field (ABF) and HAADF images were 11-23 mrad and 75-310 mrad respectively, and a collection semi-angle of 111\,mrad was used for EELS investigations. In order to improve the signal-to-noise ratio, frame series with short dwell times ($2\,\text{\textmu s/pixel}$) were used and added after cross-correlation alignment of the ABF images \cite{wang2016oxygen}. Additionally, STEM images were processed with a multivariate weighted principal component analysis routine (MSA Plugin in Digital Micrograph) to decrease the noise level \cite{Bosman06}.

Electrical resistance measurements were performed using a Quantum Design Physical Property Measurement System in van-der-Pauw geometry. The data shown in Fig.~\ref{transport} panel (a) and (b) were recorded upon slowly cooling down the sample.

In order to get insight into the differences of the electronic structure of NdNiO$_3$ grown on different crystallographic facets, we carried out DFT$+U$ calculations using the VASP (Vienna Ab initio Simulation Package) code \cite{Kresse96, Kresse99} with the generalized gradient approximation GGA-PBE functional \cite{Perdew96}. For NdNiO$_3$ an on-site Hubbard $U$ of 2\,eV was used and the in-plane lattice parameters were fixed to the values of NdGaO$_3$, while the out-of-plane lattice parameter was varied and internal atomic positions were relaxed. Note that for the [101] orientation only the $b$-axis coincides with a main in-plane crystallographic direction, while for [011] orientation it is the $a$-axis. Hence, we define the second in-plane lattice parameter as $a^* = \sqrt{a^2 + c^2}$ and $b^* = \sqrt{b^2 + c^2}$, respectively. Consequently, $a^*$ and $b$ of [101]-oriented NdNiO$_3$ were fixed to the corresponding values calculated for NdGaO$_3$, and $a$ and $b^*$ were fixed for [011]-oriented NdNiO$_3$, while in both cases the out-of-plane lattice parameters $c^*$ were varied and internal atomic positions relaxed. The parameter $c^*$ corresponds to the distance between consecutive Ni planes along the [101] and [011] direction, respectively. Since relaxations performed with the PBE functional are known to overestimate lattice constants, we compared solutions with lattice constants constrained to experimental values and found unchanged trends with variation in energy that are of the same order of magnitude and finally lead to the same conclusion. In addition we tested the possible impact of antiferromagnetic spin polarization. Since calculations become unfeasible for the large unit cells required to stabilize the experimentally observed $(\frac{1}{4} \frac{1}{4} \frac{1}{4})$ magnetic order \cite{Scagnoli08,Hepting18}, we compared the ferromagnetic results with those assuming a $G$-type $(\frac{1}{2} \frac{1}{2} \frac{1}{2})$ order and, again found negligible differences. In order to be as consistent as possible regarding relative energies, we finally used only ferromagnetic calculations for the comparisons.

\section{Experimental results}

\begin{figure}[tb]
\center\includegraphics[width=0.9\columnwidth]{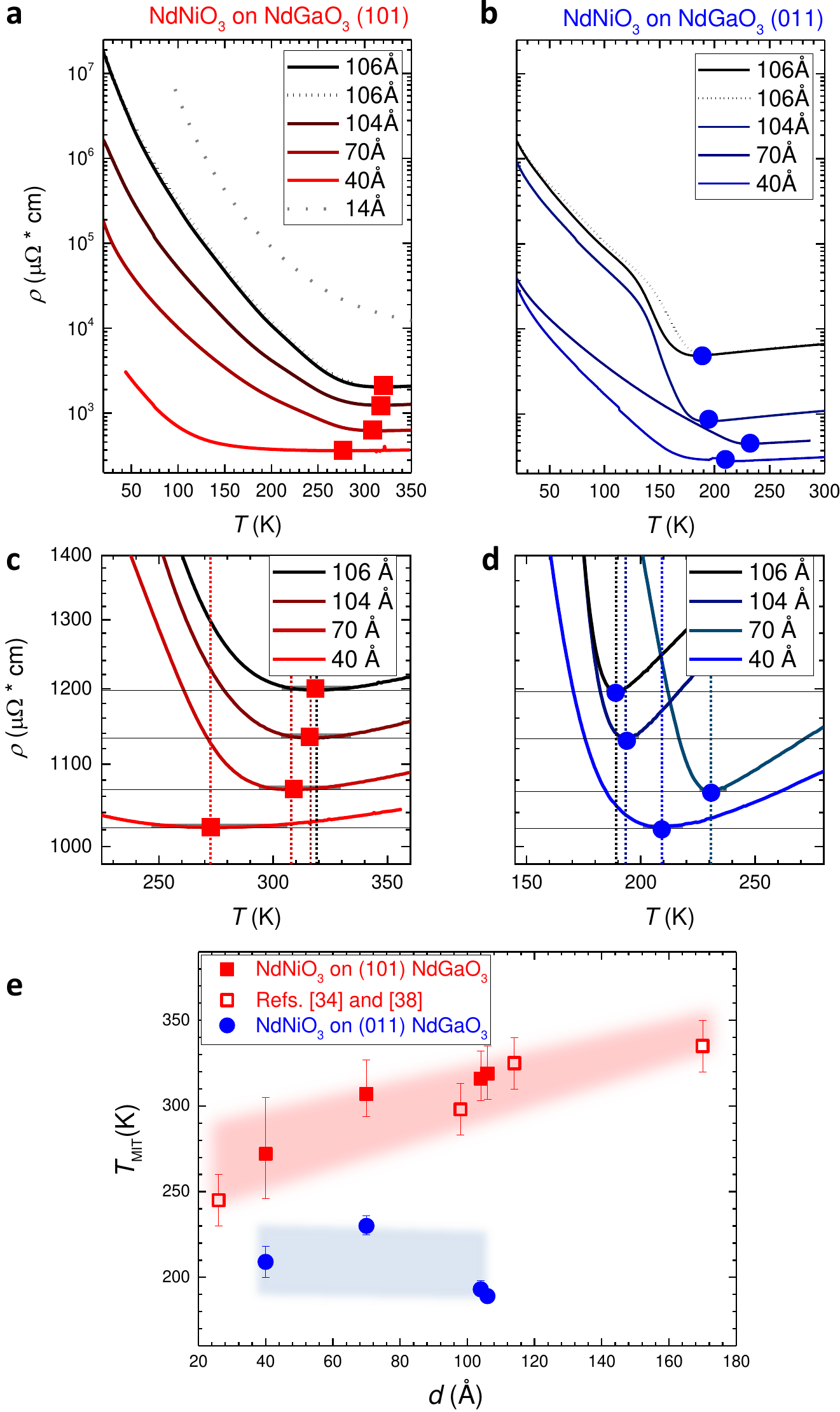}
\caption{(a), (b) Electrical resistivity $\rho$ as a function of temperature $T$ measured for NdNiO$_3$ films of different thickness grown on [101]- and [011]-oriented NdGaO$_3$ substrates, respectively. Solid lines show $\rho(T)$ measured upon cooling, while additional dashed curves show $\rho(T)$ upon warming for selected samples. The filled symbols indicate the transition temperature $T_{\text{MIT}}$, determined as described in the text. Note that the $14.4\,\text{\AA}$ film (gray dashed lines in panel a) is semiconducting without MIT. All curves, except for the one of the $14.4\,\text{\AA}$ film, are offset in vertical direction by multiples of $250\mu\Omega\cdot\text{cm}$ for clarity. (c), (d) Enlarged views of $\rho$ around the metal-insulator transition temperatures for (101) and (011) samples, respectively. Grey boxes of $\pm 5\mu\Omega\cdot\text{cm}$ hight were used to determine the error bars of $T_{MIT}$. Dotted, vertical lines mark the zero crossing of $-\partial (\text{ln}(\rho)) / \partial T$. (e) $T_{\text{MIT}}$ as a function of NdNiO$_3$ film thickness $d$. The shaded areas indicate the thickness-dependent trends, but serve only as a guide for the eye.}
\label{transport}
\end{figure}

The electrical transport properties of NdNiO$_3$ films grown on (011) and (101) NdGaO$_3$ are shown in Figs.~\ref{transport}(a, b). All films thicker than 14.4\,\AA\ show a clear change in slope in resistivity $\rho(T)$ with decreasing temperature (see the enlarged views Figs.~\ref{transport}(c, d)), signaling a transition from metallic to semiconducting behavior. Various indicators have been employed to extract the transition temperatures $T_{\text{MIT}}$ from $\rho(T)$ in previous studies on nickelates \cite{Mattoni16,Liao18,Catalano15,Dominguez19}. Here we follow Ref.~\cite{Dominguez19} and use the minimum of $\rho(T)$, i.e.\ the temperature where $-\partial (\text{ln}(\rho)) / \partial T$ changes sign, to identify the MIT. Both methods consistently give identical $T_{\text{MIT}}$ values, as indicated by the dotted, vertical lines in Fig.~\ref{transport}(c, d). Note that the absolute values of $T_{\text{MIT}}$ can vary considerably depending on the determination criterion. However, we are interested in relative trends that are reflected as long as the same criterion is applied. Within our set of samples, we find the highest value of $T_{\text{MIT}}=319$~K for the 106~\AA\ thick NdNiO$_3$ film on (101) NdGaO$_3$, which is substantially higher than the $T_{\text{MIT}}$ of 189~K of the corresponding film on (011) NdGaO$_3$, and 150~K reported for films on (001) NdGaO$_3$ \cite{Staub02}. In more detail, Fig.~\ref{transport}(e) displays the transition temperature $T_{\text{MIT}}$ as a function of NdNiO$_3$ layer thickness $d$ for both substrate orientations, with the error bars determined from the width of boxes of $\pm 5\mu\Omega$cm hight (see Fig.~\ref{transport}(c, d)). Notably, $T_{\text{MIT}}$ varies as a function of thickness $d$ (Fig.~\ref{transport}(e)). This trend is particularly pronounced in the case of [101]-oriented substrates (shown as filled, red squares in Fig. \ref{transport}(e)), and extrapolates to a $T_{\text{MIT}}$ of 335~K reported for much thicker films in Ref.~\onlinecite{Catalano15} (reproduced as open square in Fig. \ref{transport}(e)). Moreover, a comparison of the (101) and (011) films of identical nominal thickness indicates that the (101) substrate orientation generally results in a higher $T_{\text{MIT}}$ of more than 50\,K.

\begin{figure}[tb]
\center\includegraphics[width=0.99\columnwidth]{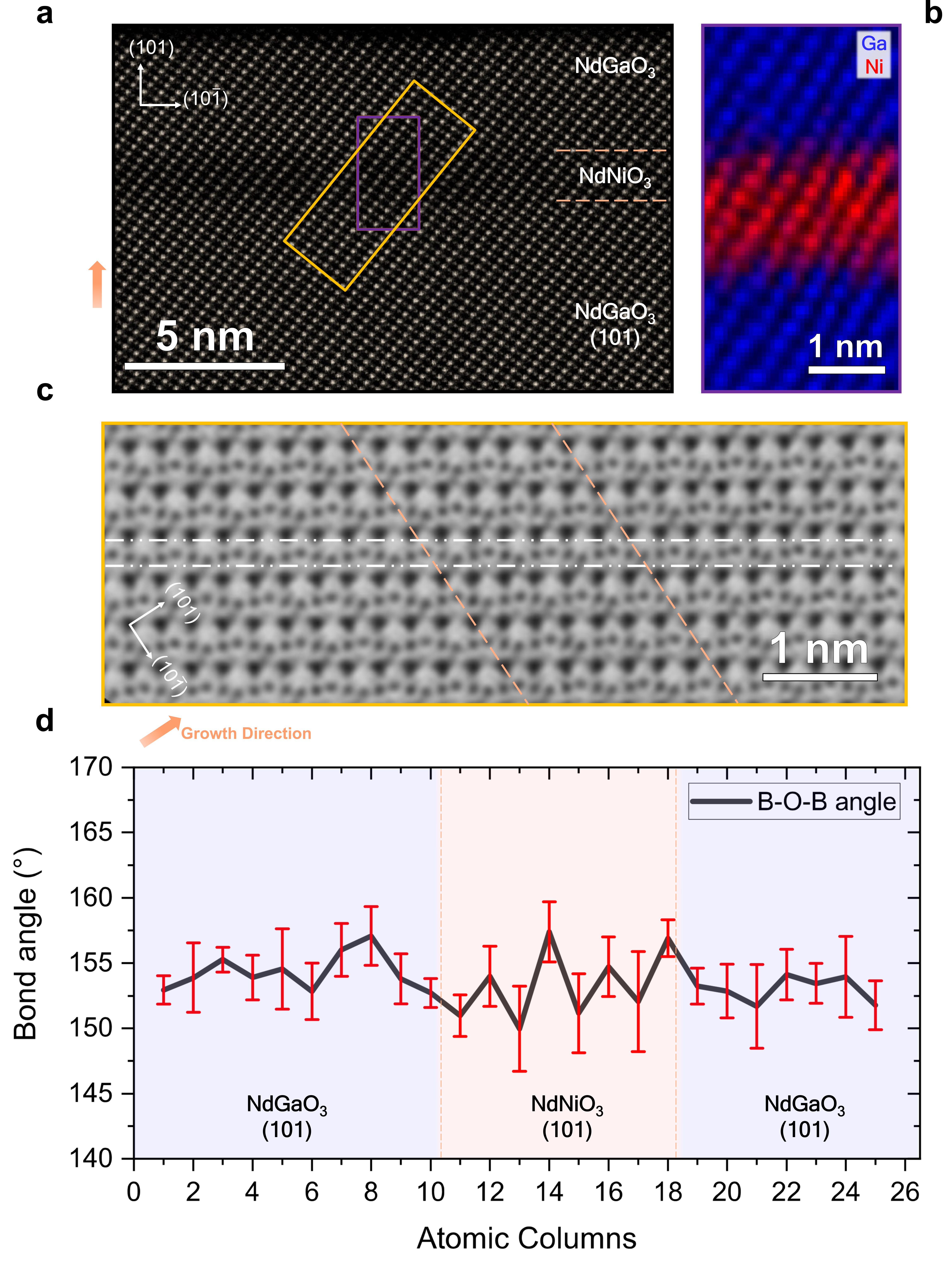}
\caption{(a) STEM-HAADF image of the ultra-thin 14.4~\AA\ NdNiO$_3$ film, including the NdGaO$_3$-NdNiO$_3$ substrate-film and the NdNiO$_3$-NdGaO$_3$ film-capping layer interfaces, indicated by dashed orange lines. Purple and orange rectangles correspond to the regions presented in panel (b) and (c), respectively. (b) Atomic-resolution two-dimensional elemental maps of Ga (blue) and Ni (red). Interfacial intermixing occurs within one monolayer at most for both interfaces. (c) The STEM-ABF image provides information about the oxygen positions. The atomic column within the region marked by the white dashed-dotted lines corresponds to \textit{B}-O-\textit{B} bonds, with \textit{B} = Ga, Ni, which are analyzed and quantified in panel (d). The dashed orange lines mark the nominal interfaces. (d) Bond angle measurement across the NdGaO$_3$-NdNiO$_3$-NdGaO$_3$ structure, indicating that the \textit{B}-O-\textit{B} bond angle remains constant within the experimental error. Error bars indicate the 95\% confidence interval, corresponding to two times the standard error. The blue and orange colored areas are guides for the eye indicating the NdGaO$_3$ and NdNiO$_3$ regions, respectively. Note that the \textit{B}-O-\textit{B} bond angle corresponds to the angle of the projection between a \textit{B}-column,  two columns of oxygen, and another \textit{B}-column, as illustrated in the corresponding projection (Fig.~\ref{structure}(b)).}
\label{8MLTEM}
\end{figure}

To gain insight into the complex facet and thickness dependence of $T_{\text{MIT}}$, we first discuss the STEM investigation of the thinnest 14.4~\AA\ sample of our series, where a MIT is absent. Note that this sample was capped by a NdGaO$_3$ layer, to avoid the influence of a free surface. The capping layer possibly has a stabilizing function on its $Pbnm$ structure. As shown in the STEM-HAADF image in Fig.~\ref{8MLTEM}(a), the 14.4~\AA\ film is of excellent crystalline quality with coherent interfaces and without traceable defects or stacking faults. Furthermore, atomic resolution STEM-EELS mapping of Ni and Ga (Fig.~\ref{8MLTEM}(b)) suggests that elemental intermixing at the interfaces is limited to one monolayer at most. Detailed information about the local octahedral distortions at the NdGaO$_3$-NdNiO$_3$ interfaces can be obtained from high-resolution STEM-ABF imaging, which allowed us to track the oxygen atomic column positions (Fig.~\ref{8MLTEM}(c)). The analysis of the ABF image is shown in Fig.~\ref{8MLTEM}(d), quantifying the bond angle between the $B$-site cation (Ga, Ni) and oxygen from the substrate region across the film and into the capping layer. Importantly, we find that across the interfaces the Ga-O-Ga and Ni-O-Ni bond angles remain constant within the experimental error, suggesting that the oxygen positions of the NdNiO$_3$ layer are forced to the oxygen positions given by NdGaO$_3$. In other words, this implies that the Ga-O-Ga bond angle of NdGaO$_3$ is transferred to the Ni-O-Ni angle in the epitaxial NdNiO$_3$ layer, which is consistent with a bond-angle transfer scenario from NdGaO$_3$ substrates to NdNiO$_3$ films proposed in Ref.~\onlinecite{Catalano15}. For nickelate films thicker than 14.4~\AA , however, the ABF analysis in Ref.~\onlinecite{Hepting18} revealed a subsequent relaxation of the Ni-O-Ni bond angles up to bulk-like values, indicating that the strongly enhanced $T_{\text{MIT}}$ of thicker NdNiO$_3$ films on [101]-oriented NdGaO$_3$ substrates is likely not caused by altered Ni-O-Ni bond angles, but by effects that will be discussed later. When lowering the temperature, the pinning of oxygen positions in the vicinity of the film-substrate interface obstructs the NiO$_6$ octahedral breathing-mode distortion, and suppresses the MIT. As a result, in thicker films, a thin NdNiO$_3$ layer close to the interface of the NdGaO$_3$ substrate shows no bond-disproportionation down to lowest temperatures \cite{Hepting18}, while layers further away from the interface gradually develop a breathing distortion, which sets the basis for the insulating state. In this scenario, and within the range of film thicknesses considered in Fig.~\ref{transport}, $T_{\text{MIT}}$ naturally decreases as a function of reduced film thickness, with a crossover to a regime of ultra-thin films, where the octahedral breathing mode is fully suppressed.

 \begin{figure}[tb]
 \center\includegraphics[width=0.99\columnwidth]{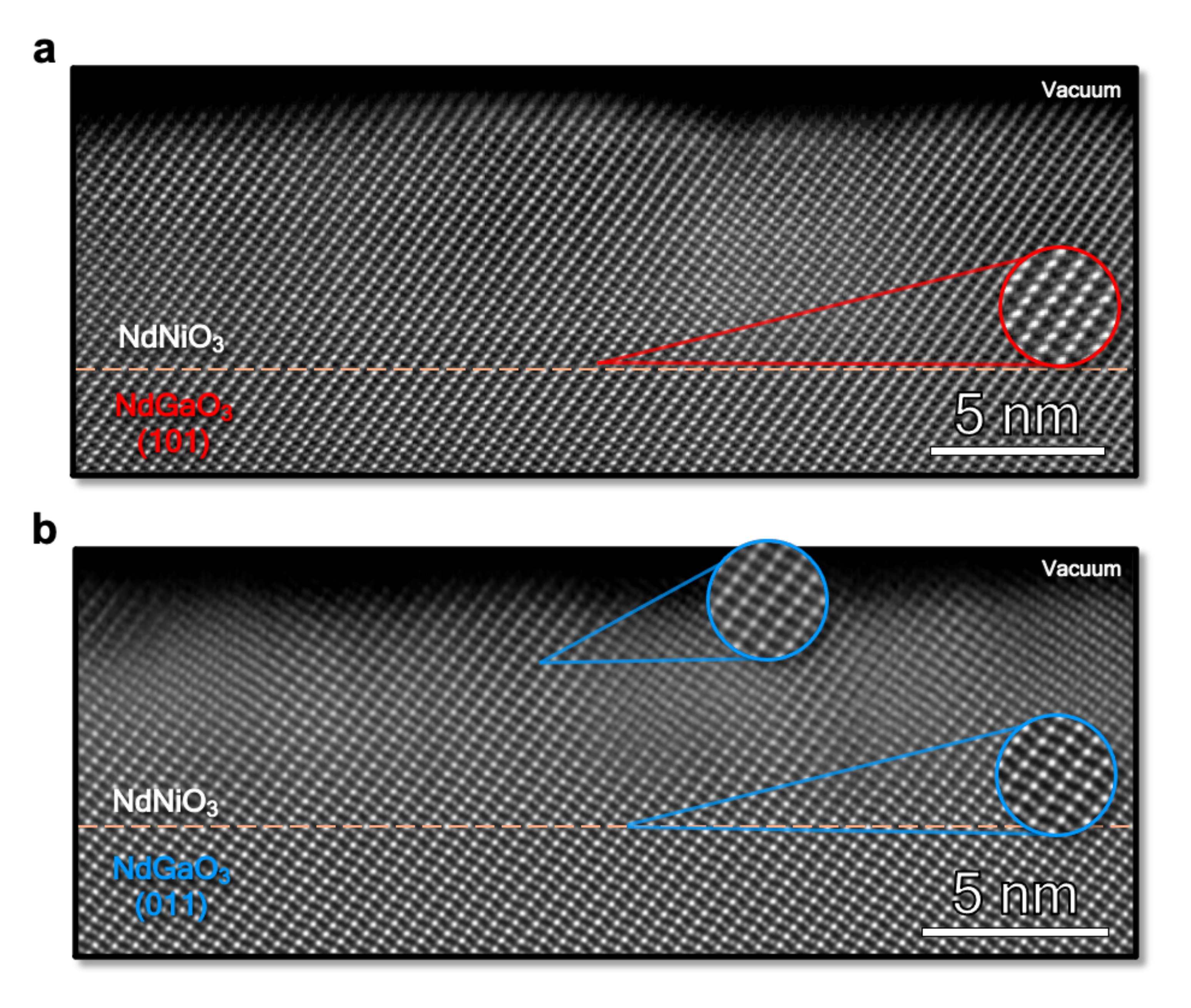}
 \caption{(a), (b) STEM-HAADF images of the 70~\AA\ NdNiO$_3$ films on [101]- and [011]-oriented NdGaO$_3$ substrates, respectively. The insets illustrate the presence of characteristic \textit{straight} and \textit{zig-zag} patterns of the Ni and Nd cation positions, which are indicative of the [101] and [011] orientations, respectively. While the NdNiO$_3$ film on (101) NdGaO$_3$ is entirely [101]-oriented and of high crystalline quality (a), a more defective structure together with reoriented (101) patches is observed for the film on (011) NdGaO$_3$.}
 	\label{TEM_KF}
 \end{figure}

Next, we clarify the mechanisms that are responsible for the different $T_{\text{MIT}}$ of films with nominally identical thickness, but grown on different facets. To this end, we performed STEM-HAADF imaging with atomic resolution on the two 70~\AA -thick films. A representative HAADF image of each sample is shown in Fig.~\ref{TEM_KF}. For [101] substrate orientation, we find that the NdNiO$_3$ exhibits high epitaxial quality (Fig.~\ref{TEM_KF}(a)). In particular, the Nd and Ga ions in the NdGaO$_3$ substrate are arranged in \textit{straight lines}, as expected for [101]-oriented films (see also Fig.~\ref{structure}(b)). A closely similar \textit{straight line} pattern is adopted by the Nd and Ni ions in the film and persists up to the topmost layers, suggesting that the entire film grew in the same orientation as the substrate. By contrast, several regions with crystal defects can be observed in the film on [011]-oriented NdGaO$_3$ (Fig.~\ref{TEM_KF}(b)), which was grown simultaneously under identical conditions (see the Methods section). While the first monolayers of the film exhibit good crystalline quality, the defective regions start to form a few monolayers after the film-substrate interface. The majority of defects in Fig.~\ref{TEM_KF}(b) can be attributed to stacking faults of the perovskite crystal lattice. However, a close inspection of the upper parts between defective regions reveals some patches with rather \textit{straight lines} of Nd and Ni, while overall the Nd and Ni ions are mostly arranged in \textit{zig-zag lines} in accordance with the [011] orientation. Remarkably, this change of the Nd and Ni arrangement is indicative of a reorientation of the NdNiO$_3$ unit cell with respect to the growth direction, \textit{i.e.} from [011] to [101] orientation.
The higher defect density and the presence of partially reoriented areas in (011) films indicate an opposite phase preference resulting from lattice mismatch and the need for interfacial connectivity. To verify and quantify these possibly competing energy scales, we consider different epitaxial combinations between film and substrate orientations in a strain minimization scenario, and compare their ground state energies in DFT$+U$ calculations.

\section{DFT$+U$ calculations}

 \begin{figure*}[t]
 \center\includegraphics[width=0.85\textwidth]{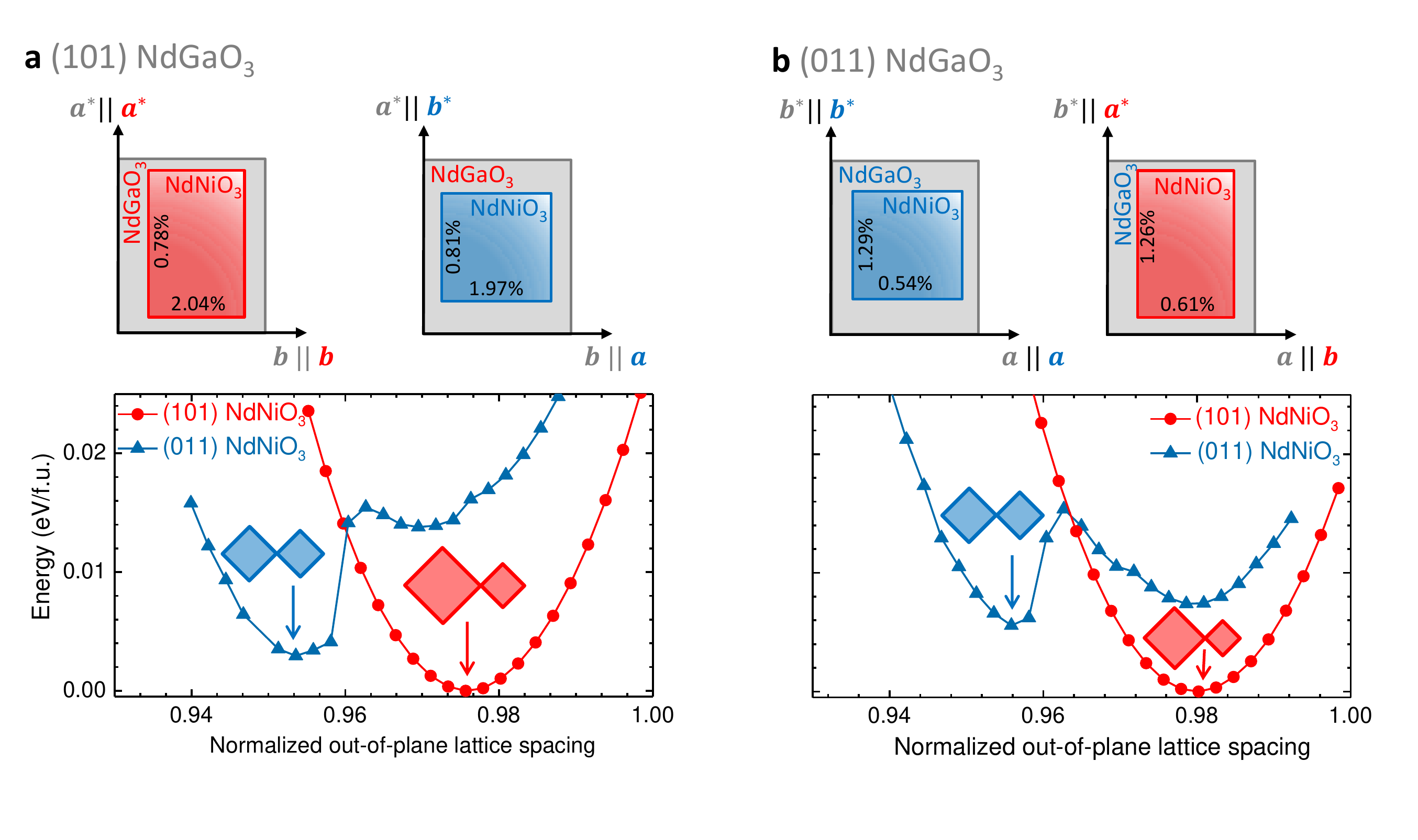}
 \caption{Top panels: Schematic projections of the dimensions of the bulk NdNiO$_3$ and NdGaO$_3$ unit cells for different film orientations on (a) (101) NdGaO$_3$ and (b) (011) NdGaO$_3$ substrates. The crystallographic in-plane directions ($a^*$, $b^*$, $a$, $b$) of NdNiO$_3$ and NdGaO$_3$ are indicated, together with the normalized, directional lattice mismatch $(l_{\rm NdGaO_3}-l_{\rm NdNiO_3})/l_{\rm NdNiO_3}$ in $\%$. Bottom panels: Ground state energy of NdNiO$_3$ as a function of the normalized out-of-plane lattice spacing as obtained from DFT$+U$ calculations for the four different substrate-film orientation configurations depicted in the top panels. Keeping the in-plane lattice parameters fixed to those of [101]-oriented NdGaO$_3$ in (a) and [011]-oriented NdGaO$_3$ in (b), the ground state energies have been calculated for a range of values for the out-of-plane lattice spacing. Since the NdGaO$_3$ substrate induces tensile strain in all configurations, a reduced $c^*$ spacing is expected with respect to the bulk (for the normalization we used bulk NdNiO$_3$\cite{Garica-Munoz09}, that is $c^*_{(101)}=4.397$\,\AA\ for the (101) phase and $c^*_{(011)}=4.394$\,\AA\ for the (011) phase). Furthermore, the DFT$+U$ derived structures reveal an enhanced bond-disproportionation in [101]- compared to [011]-oriented films on both substrate orientations, as illustrated by the insets.}
 	\label{DFT}
 \end{figure*}

In order to understand the observed differences in $T_{\text{MIT}}$ of films grown on the different NdGaO$_3$ facets, we first address the lattice mismatch as origin of preferred film orientation. Determined by the choice of the substrate facet, NdNiO$_3$ experiences a slightly different lattice mismatch depending on its own relative orientation. This is illustrated in the top part of Fig.~\ref{DFT}(a,b). For all possible combinations of film and substrate orientation, the substrate induces tensile strain to the NdNiO$_3$ film. For the one-to-one configurations we find $\approx0.8\%$ and $\approx2.0\%$ in-plane lattice mismatch for (101)-(101) NdNiO$_3$-NdGaO$_3$ (Fig.~\ref{DFT}(a)) and $\approx1.3\%$ and $\approx0.5\%$ for (011)-(011) (Fig.~\ref{DFT}(b)). The values for the other two arguably counter-intuitive configurations of (011)-(101) and (101)-(011), however, do not differ significantly. Thus, strain minimization alone cannot explain the experimentally observed partial reorientation to (101) for films grown on (011) substrates (Fig.~\ref{TEM_KF}(b)).

Next we address the relevance of electronic interactions on the level of DFT$+U$ by discussing the results obtained for the four configurations illustrated in the top panels of Fig.~\ref{DFT} using a Hubbard $U$ of $2\,\text{eV}$, typical for rare-earth nickelates \cite{Hampel17, Varignon17}. The bottom panels of Fig.~\ref{DFT}(a,b) show the ground state energy of NdNiO$_3$ as a function of the normalized out-of-plane lattice spacing $c^*_{(101)}/c_{(101)}^{* \rm bulk}$ and $c^*_{(011)}/c_{(011)}^{*\rm bulk}$, respectively (bulk values were taken from Ref.~\onlinecite{Garica-Munoz09}), while the in-plane lattice constants were fixed to those of (011) and (101) NdGaO$_3$, respectively. On both substrate facets the growth of [011]-oriented NdNiO$_3$ has an overall higher energy and is therefore less favored (bottom panels in Fig.~\ref{DFT}(a,b)). For [101]-oriented NdNiO$_3$ we find a global minimum on both substrate facets for reduced out-of-plane lattice spacings close to 4.31~\AA, in good agreement with out-of-plane lattice spacing of (101) NdNiO$_3$ films determined by resonant x-ray reflectivity \cite{Hepting18}. Thus we conclude that the combination of strain and electronic interactions captured in our DFT$+U$ calculations indicate a preference of [101] film orientation independent of the substrate facet. In addition the DFT$+U$ calculations reveal a second important result: the crystal structures corresponding to the global energy minima differ significantly from the bulk with respect to the size of the bond-disproportionation. Specifically, in the bond-disproportionated ground state of (101) NdNiO$_3$, the long ($d_{\text{L}}$) and short ($d_{\text{S}}$) Ni-O bond lengths in the expanded and compressed NiO$_6$ octahedra are $d_{\text{L}} = 1.97$~\AA\ and $d_{\text{S}} = 1.90$~\AA, respectively. Parametrization of the distortion $\delta d$ as the difference in long and short bond lengths $d_{\text{L}}$ and $d_{\text{S}}$ from the mean value, yields $\delta d = 0.035$~\AA, which is significantly enhanced compared to a $\delta d$ of 0.015~\AA\ calculated for a fully relaxed bulk NdNiO$_3$ unit cell \footnote{The exact numeric value of $\delta d$ depends on details of the calculation,
such as the choice of $U$. However, we emphasize that the trend of an enhanced bond-disproportionation for (101) NdNiO$_3$ is robust.}. Conversely, the distortion $\delta d$ of the hypothetical, local minimum structures of (011) NdNiO$_3$, either on (101) or (011) NdGaO$_3$ is similar to that of bulk NdNiO$_3$ \footnote{Importantly, the crystal structures of (011) NdNiO$_3$ (all data points in Fig.~\ref{DFT}, including the side-minima) do not exhibit enhanced bond-disproportionation}. Therefore, we conclude that the orientational structure stabilization is not driven by strain minimization alone, but by its combination with electronic interactions. In other words, the system chooses the orientation with a particular tensile lattice mismatch that maximizes the bond-disproportionation.

This explains well our observations for thick films on [101]-oriented substrates, where we clearly observe the stabilization of (101) film orientation (STEM image in Fig.~\ref{TEM_KF}) and an enhancement of $T_{\text{MIT}}$ compared to the bulk. However our DFT$+U$ calculations do not account for effects of interconnectivity in the vicinity of the film-substrate interface. Due to the necessity to form Ga-O-Ni bonds across the interface and the stability of the NdGaO$_3$ $Pbnm$ structure (with only one Ni Wyckoff position $4a$) down to the lowest temperature, this interfacial connectivity prevents the development of bond-disproportionation close to the interface. In the ABF image analysis of Fig.~\ref{8MLTEM} we found that the length scale, on which the oxygen atomic positions are pinned by the [101]-oriented substrate, is of the order of 14~\AA\ (8 atomic layers). Similarly, the HAADF image in Fig.~\ref{TEM_KF}(b) shows that for the (011) film growth the first 5-10 atomic layers of defect-free [011]-oriented NdNiO$_3$ are a related length scale. Above the critical length scale, where the pinning effect of the substrate is relaxed, the above discussed combination of lattice mismatch and electronic correlations takes over and causes the partial reorientation to (101) film alignment and defect formation as observed in the top part of the STEM image in Fig.~\ref{TEM_KF}(b).

We conclude that closely similar energy scales of lattice and electronic degrees of freedom in rare-earth nickelates enable the heteroepitaxial stabilization of phases with enhanced electronic interactions. In particular, our study shows that the symmetry of the substrate as well as the choice of the crystal facet can have significant impact on the electronic properties of NdNiO$_3$, as demonstrated by the complex thickness and facet dependence of $T_{\text{MIT}}$. Up to thicknesses of 5-10 atomic layers, the structural pinning of the substrates dominates and determines the orientation and suppression of bond-disproportionation in the NdNiO$_3$ slabs. Above this length scale, lattice mismatch in combination with electronic interactions determine the properties of the films. Our study therefore provides an extraordinarily high level of understanding and control of the bond-disproportionated phase in NdNiO$_3$.

\section{Summary}

In summary, we investigated in detail the heteroepitaxial modifications in NdNiO$_3$ thin films by combining analytical high-resolution STEM, electrical transport measurements, and DFT$+U$ calculations. We identified a complex interplay between lattice constraints and electronic interactions, which manifests itself in a complicated dependence of the MIT temperature with film thickness and substrate orientation. Finally, we established a picture of competing energy scales that facilitates the stabilization of phases with modified bond-disproportionation by means of facet and thickness control in heterostructures. The amended bond-disproportionation has important implications for the magnetic ground state and the exchange interactions, which can be qualitatively different from the bulk \cite{Hepting18,Lu18,Fuersich19}.

The controlled preparation of phases opens a new route to tune material properties in heterostructures for future applications in functional devices.
This type of heteroepitaxial control could be used to stabilize exotic phases in $R$NiO$_3$ such as multiferroicity \cite{vandenBrink08, Giovannetti09} and superconductivity \cite{Chaloupka08,Hansmann09,Li2019} and to make them accessible in a technologically relevant parameter range. It would also be interesting to further study the connection of structural effects (and the strength of bond-order) to the emergence of magnetic order. This could possibly be done by using other orthorhombic substrates to tune the nature and the size of the strain.

\begin{acknowledgments}
We acknowledge U.~Salzberger for TEM specimen preparation and B.~Lemke for technical support. Y.E.S. and P.v.A. acknowledge financial support from the European Union's Horizon 2020 research and innovation program under grant agreement No. 823717 (ESTEEM3). K.F., B.K., and E.B. acknowledge financial support by the Deutsche Forschungsgemeinschaft (DFG, project no. 107745057, TRR80).

Y.E.S.\ and K.F.\ contributed equally to this work.
\end{acknowledgments}


%

\end{document}